\documentstyle[11pt,aaspp4]{article}
\begin{document}
\title{Delayed GeV--TeV Photons from Gamma-Ray Bursts Producing High-Energy 
Cosmic Rays}
\author{Eli Waxman}
\affil{Institute for Advanced Study, Princeton, NJ 08540}
\authoremail{waxman@sns.ias.edu}
\author{Paolo Coppi}
\affil{Department of Astronomy, Yale University, P.O. 
Box 208101, New Haven, CT 06520-8101}
\authoremail{coppi@astro.yale.edu}

\begin{abstract}

A scenario in which cosmic rays (CRs) above $10^{20}{\rm eV}$ are produced 
by cosmological gamma-ray bursts (GRBs) is consistent with observations 
provided that deflections by the inter-galactic magnetic field (IGMF)
delay and spread the arrival time of the CRs over $\geq50{\rm yr}$. 
The energy lost by the CRs as they propagate and interact with the 
microwave background is transformed by cascading into secondary GeV-TeV 
photons. We show that a significant fraction of these photons can arrive with 
delays much smaller than the CR delay if much of inter-galactic space is
occupied by large-scale magnetic ``voids'', regions of size
$\gtrsim5{\rm Mpc}$ and field weaker than $10^{-15}{\rm G}$. 
Such voids might be expected, for example, in models where a weak primordial 
field is amplified in shocked, turbulent regions of the intergalactic medium
during the formation of large-scale structure.
For a field strength $\sim4\times10^{-11}{\rm G}$ in the high field regions, 
the value required to account for observed galactic fields if the IGMF were 
frozen in the protogalactic plasma, the delay of CRs produced by a burst at a 
distance of $100{\rm Mpc}$ is $\sim100{\rm yr}$, and the fluence of secondary 
photons above $10{\rm GeV}$ on hour--day time scales is 
$I(>E)\sim10^{-6}E_{\rm TeV}^{-1}{\rm cm}^{-2}$. This fluence is 
close to the detection threshold of current high-energy $\gamma$-ray 
experiments. Detection of the delayed flux would support the GRB-CR 
association and would also provide information on the IGMF structure.

\end{abstract}

\section{Introduction}

Recent gamma ray and cosmic ray observations give increasing evidence
that the sources of gamma ray bursts (GRBs) and of cosmic rays (CRs) with 
energy $E>10^{19}{\rm eV}$ are cosmological (see \cite{cos} 
for GRB observations review; \cite{Fly1}, \cite{AGASA2}, \cite{Wb} for CRs).
The sources of both phenomena, however, remain unknown. In particular, 
most of the CR sources discussed so far have difficulties in accelerating
CRs up to the highest observed energies (e.g., \cite{huge}).
Although the source of GRBs is unknown, their observational
characteristics impose strong constraints on the physical conditions
in the $\gamma$-ray emitting region (\cite{scen1}, \cite{scen2}),
which imply that protons may be accelerated in this 
region to energies $10^{20} - 10^{21} {\rm eV}$ (\cite{Wa}, \cite{Vietri}).
In addition, the average rate (over volume and time) at which energy
is emitted as $\gamma$-rays by GRBs and in CRs above $10^{19}
{\rm eV}$ in the cosmological scenario is, remarkably, comparable 
(\cite{Wa},b). These two facts suggest that GRBs and high-energy CRs may 
have a common origin.

  An essential ingredient of the GRB model for CRs is the
time delay due to intergalactic magnetic fields. The energy of the most
energetic CR detected by the Fly's Eye experiment is in excess of
$2\times10^{20}{\rm eV}$ (\cite{Fly1}), and that of the most
energetic AGASA event is above $10^{20}{\rm eV}$ (\cite{AGASA2}). On a
cosmological scale, the distance traveled by such energetic particles is
small: $<100{\rm Mpc}$ ($50{\rm Mpc}$) for the AGASA (Fly's Eye) event
(e.g., \cite{huge}). Thus, the detection of these events over a $\sim5
{\rm yr}$ period can be reconciled with the rate of nearby GRBs, $\sim1$
per $50\, {\rm yr}$ in the field of view of the CR experiments out to $100
{\rm Mpc}$ in a standard cosmological scenario (e.g., \cite{rate2}), only if
there is a large dispersion, $\geq50{\rm yr}$, in the arrival time of protons 
produced in a single burst (Note, that this implies that if a direct 
correlation between 
high energy CR events and GRBs, as recently suggested by \cite{MU}, is observed
on a $\sim10{\rm yr}$ time scale, it would be strong evidence {\it against} a 
cosmological GRB hypothesis). The required dispersion may result from 
deflections of CR protons by the inter-galactic magnetic field 
(\cite{Wa}).

The inter-galactic magnetic field (IGMF) has not been detected so far. 
Faraday-rotation measures set an upper limit of $\sim10^{-9}{\rm G}$ for 
a field with $1 {\rm Mpc}$ correlation length (see \cite{Kron} for review). 
Other methods have recently been proposed to probe fields in the range
$10^{-10}$--$10^{-20}{\rm G}$ (e.g., \cite{Plaga}, \cite{Olinto}, \cite{Avi}).
Theoretical considerations regarding the existence and strength of the IGMF 
are related to the formation of the observed $\mu{\rm G}$ fields in galaxies. 
Recent studies suggest that a galactic dynamo cannot produce
the observed large-scale fields in galactic disks (\cite{Anderson})
and that one must turn to alternative mechanisms, which
typically rely on a pre-existing field.
Galactic fields might be created, for example, by
compression of much weaker fields in collapsing protogalactic regions.
This mechanism requires a protogalactic field of strength
$10^{-11}$--$10^{-10}{\rm G}$ and correlation length of order
$1{\rm Mpc}$. Such fields could be primordial, in which case they would
likely permeate all intergalactic space. However, this need not be the
case. For example, such fields could be generated from
a much weaker primordial field, $\sim 10^{-20}$~G,
due to the turbulence induced in the formation of large scale
structure in the universe (\cite{Kulsrud}). 
In this picture, the IGMF would ``trace the mass'', with high
$10^{-11}$--$10^{-10}{\rm G}$ fields in the high density (proto-galactic)
regions of the large scale structure and very low fields in the 
intervening voids.

Most of the energy lost by the CRs as they propagate and interact with the 
microwave background is transformed by cascading into secondary GeV-TeV 
photons (e.g., \cite{cas3}, \cite{cas1}). In this {\it Letter} we
show that even though the CR time delay must be $\gtrsim50{\rm yr}$, 
a significant fraction of 
the GeV-TeV cascade radiation can arrive with much shorter
delays, on the order of hours to days,
provided that a large fraction of the inter-galactic space is occupied
by magnetic ``voids'', regions of very low magnetic field ($<10^{-15}$ G).
In \S 2 we present a qualitative discussion of the development
of electro-magnetic cascades in the presence of an IGMF. The expected
high energy photon flux is calculated using detailed Monte-Carlo simulations 
in \S 3. Implications for current and future 
high energy gamma-ray experiments are discussed in \S 4.

\section{Electro-magnetic Cascades and the Inter-galactic Magnetic Field}

As they propagate, nucleons with energy in excess of $10^{20}{\rm eV}$ 
lose energy due to the photo-production of pions in interactions 
with microwave background photons.
Pion decay converts $\sim40\%$ of the energy lost by nucleons 
to neutrinos, and the rest to photons, electrons and positrons. 
The energetic, $\sim10^{19}{\rm eV}$,
secondary photons, electrons and positrons further interact with 
the microwave background to form electromagnetic cascades:
high energy photons interact with background photons and
produce electron-positron pairs, which, in turn, lose their energy by
inverse-Compton scattering of background photons. The mean energy of the 
secondary photons 
is degraded until it drops below the threshold for pair 
production, $\sim10^{14}{\rm eV}$ for interaction with microwave photons. 
The development of an electro-magnetic cascade, typically
over a $\sim10{\rm Mpc}$ distance, converts
most of the secondary particles energy to $\sim{\rm TeV}$ photons. 
The distance out to which $>{\rm TeV}$ cascade photons may be
observed is limited since they may further interact with 
infrared background photons. Current limits 
on the infrared background radiation energy density 
constrain the mean free path for pair production 
on infrared photons to lie in the range $\sim 0.6-2{\rm Gpc}$ for
$1{\rm TeV}$ photons and  $\sim 20-200{\rm Mpc}$ for $10-100{\rm TeV}$ photons
(\cite{Stecker}).

We now estimate the arrival time delays of cascade photons resulting
from deflections by the IGMF of the protons initiating the cascades and of 
the pairs produced during the cascades. We consider the propagation 
of protons and secondary particles through an IGMF where a fraction 
$\eta$ of intergalactic space is occupied by structures of typical size 
$\lambda$ and coherent magnetic field $B$, and where the rest of space is 
filled with a much weaker field. 

First, consider the deflection of a proton with energy
$E_p.$ After traveling a distance $d$ through the IGMF, the proton
accumulates a typical deflection angle 
$\theta_p\sim(\eta d/\lambda)^{1/2}\lambda/R_L$,
where $R_L=E_p/eB$ is the Larmor radius.
This deflection results in a time delay 
\begin{equation}
\tau\sim \theta_p^2d/2c\approx{2\over3}\left({eB\over E_p}d\right)^2
{\lambda \eta \over 2c}\simeq60
\left({B_{-11}\over E_{p,20}}d_{100}\right)^2\eta_{0.2}\lambda_{3}
{\rm\ yr},
\label{delay}
\end{equation}
where $E_p=10^{20}E_{p,20}{\rm eV}$, $d=100d_{100}{\rm Mpc}$,$\eta=0.2
\eta_{0.2}$, $B=10^{-11}B_{-11}{\rm G}$ and $\lambda=3\lambda_{3}{\rm Mpc}$ 
(the $2/3$ factor is due to random field orientations).
For IGMF parameters that might be expected in models where galactic fields
result from the compression of an IGMF produced by turbulence induced in the 
formation of large-scale structure, $\eta\sim0.2$, $\lambda
\sim1{\rm Mpc}$ and $B\sim4\times10^{-11}{\rm G}$, the time delay accumulated 
by CR protons propagating $\sim100{\rm Mpc}$ distance is large enough to 
reconcile the observed CR event rate with the rate of nearby GRBs 
($\tau$ depends sensitively on $E_p$, so 
that the stochastic proton energy loss via pion production results in
a broadening of the CR pulse over a time comparable to $\tau$.)
The time delay for cascade photons may be much shorter than the CR delay, 
since the protons lose a significant fraction of their energy over a distance 
much shorter than the $\sim100{\rm Mpc}$ distance
over which they accumulate their total time delay.
Protons of initial 
energy $>2\times10^{20}{\rm eV}$ lose $10\%$ of their energy,
and therefore initiate electro-magnetic cascades carrying $\sim 10\%$ of the 
total cascade radiation energy, over a distance 
$d_{\rm init}\sim1{\rm Mpc}$. 
If the secondary photons and subsequent cascade particles produced in these
cascades were to propagate rectilinearly to the observer, 
the time delay of the resulting cascade photons would be 
$\leq d_{init}^3/6R_L^2c\sim0.2B_{-11}^2{\rm day}$ if the burst went off in a 
high field region, or much shorter if the burst happened to go off in a 
region of low field. If 
bursts occur inside galaxies, regions of potentially much higher field ($\mu
{\rm G}$ on a kpc scale), the initial time delay accumulated by $3\times10^{20}
{\rm eV}$ protons as they traverse and leave the host galaxy is only $\sim 
1$ day.

The pairs produced in a cascade are also subject to deflection
by the IGMF, and this can lead to much larger photon arrival time delays.
The Compton energy loss distance for pairs of energy 
$E_e=10^{18},\ 10^{15}{\rm eV}$ 
(corresponding to the high, low energy ends of the cascade), is $\Lambda_{IC}
\simeq1,\ 0.01{\rm Mpc}$.
If these pairs were created in a high field region, their deflection by
the magnetic field, $\theta_e\sim eB\Lambda_{IC}/E_e$, would delay
subsequent cascade photons by $\sim10^3,\ 10^5B_{-11}^2{\rm yr}$. 
Thus, the only photons arriving with reasonably short time delays
are those produced by branches of the cascades where
pairs were produced in IGMF voids. (A minimum delay shorter than
day requires a ``void'' field $<10^{-15}{\rm G}$). The following
characteristics of electro-magnetic cascades development ensure, that
a significant fraction of the cascade photons are produced by such branches,
provided that a large fraction of the inter-galactic space is occupied by 
large scale, $\sim5{\rm Mpc}$, voids.
The distance over which CR protons decelerate, $\sim10{\rm Mpc}$, and
the mean free paths of photons at the high-energy ($10^{18}-10^{19}{\rm eV}$) 
part of the cascade, $\sim1{\rm Mpc}$, are comparable to or larger than
the typical size of the high field regions. This ensures that a significant 
fraction of the cascades are initiated inside voids, and also implies
that, at the high energy end of the cascade, a particular cascade branch 
can ``skip'' over high field regions without accumulating time delay.
The short mean free paths of particles at the low energy part of the
cascade, $\Lambda_{IC}<0.1{\rm Mpc}$ for $E_e<10^{17}{\rm eV}$, then
imply that the development of a branch, whose $\sim10^{17}{\rm eV}$ 
particles are created in a void region, is likely to be completed in 
this region.

\section{Numerical Results}

We now present numerical results for the flux of delayed cascade 
photons  expected in an IGMF where a large fraction of inter-galactic 
space is occupied by large-scale voids, and the field strength in the high 
field regions is high enough to produce a $\sim100{\rm yr}$
delay in the CR arrival time. We constructed a Monte Carlo code 
that propagates high energy nucleons
through the microwave background and calculates the evolution of the 
cascades they initiate. All the relevant 
particle interaction processes are included, 
and the gyration of charged particles about the IGMF 
is explicitly calculated. Since we are only interested in source distances 
of order $200{\rm Mpc}$, cosmological evolution effects were not included.
The cosmic background radiation field was chosen to be a
superposition of: (i) a blackbody of temperature $2.7\arcdeg{\rm K}$; 
(ii) an infrared/optical power law distribution at energies 
$\epsilon>0.02{\rm eV}$ with density $n(\epsilon) = 8\times 10^{-4}
(\epsilon/{\rm eV})^{-2}\, {\rm cm}^{-3} {\rm eV}^{-1}$, corresponding to
the lower bound of current estimates of the intensity; 
(iii) a truncated power law distribution at radio energies 
with density $n(\epsilon) = 3.5\times10^{-7} (\epsilon/{\rm eV})^{-1.75}
\exp\bigl [-\epsilon/10^{-5}{\rm eV}] {\rm cm}^{-3} {\rm eV}^{-1}$ 
(e.g., \cite{Sironi}). Our results depend only weakly on the radio background 
intensity, since the cascades are initiated by particles with
energies $\lesssim10^{19}{\rm eV}$, for which interaction with the radio 
background is only marginally important.
The infrared/optical
radiation does not influence significantly the $<10{\rm TeV}$
flux, but may strongly affect the flux at higher energies. 
This is further addressed below.

We generated several IGMF structures by sampling the sizes of high(low)
field regions along the line of sight from exponential size distribution
with mean $2{\rm Mpc}$($10{\rm Mpc}$). 
The field in each region was randomly oriented with magnitude 
$B=4\times 10^{-11}{\rm G}$ in the high field regions and $10^{-20}{\rm G}$
in the ``void'' regions (the ``void'' field can be as high as 
$\sim10^{-16}{\rm G}$ without significantly affecting the results
shown).  The resulting distributions for $\geq10^{20}{\rm eV}$
CR arrival time delays are shown in Fig. 1 for
five different realizations of the IGMF. The source was chosen to be at a 
distance of $100{\rm Mpc}$, and assumed to produce a power law 
proton spectrum $dN/dE \propto E^{-2}$ in the range $10^{20}$--
$10^{21}{\rm eV}$.
The corresponding photon fluences above $1{\rm TeV}$ are shown in Fig. 2. The 
fluence is normalized for a burst producing $10^{51}{\rm ergs}$ in
$>10^{20}{\rm eV}$ protons, as would be required to account for
the observed CR flux.
The cascade radiation spectrum is determined by the development of
the low energy end of the cascade, and is therefore independent of the
proton production spectrum. For the short time delays of interest
here, the low energy development of the relevant cascade branches
always occurs in void regions. Thus, the spectrum is similar
to that obtained in the absence of an IGMF and is time independent. 
The integral  photon number
flux (above energy $E$) is approximately given by
$I(>E) \propto E^{-0.8}$ for $3{\rm GeV}<E<200{\rm GeV}$,
$I(>E) \propto E^{-1.0}$ for $200{\rm GeV}<E<10{\rm TeV}$,
and $I(>E) \propto E^{-1.4}$ for $10{\rm TeV}<E<{70 TeV}$. 

The spread in the typical CR time delay obtained for different field
structure realizations is small, since this delay is accumulated over
a $100{\rm Mpc}$ propagation. The spread in the photon fluence obtained 
on a time scale of days is larger, since it is determined by
the IGMF structure near the source, at distances $\leq20{\rm Mpc}$, where
the development of most of the relevant electro-magnetic cascades take
place: the highest fluence was obtained for a case where the first 
$20{\rm Mpc}$ were free of high field regions, and the lowest for a case where
$50\%$ of the first $20{\rm Mpc}$ were occupied by two $\sim5{\rm Mpc}$ high 
field regions. This reflects the results of a large set of Monte-Carlo
calculations, which show that for $B\sim4\times10^{-11}{\rm G}$ and
void regions size in the range $5$--$10{\rm Mpc}$, the fluence increases from
$\sim10^{-7}{\rm cm}^{-2}$ to $\sim10^{-6}{\rm cm}^{-2}$ as the fraction of the
volume occupied by high field regions within $20{\rm Mpc}$ from the source
decreases from $0.4$ to $0.2$. The presence of a host
galaxy near the burst is expected to introduce a minimum delay of
order $1{\rm day}$ in the cascade photon arrival times, 
and is not expected to
significantly affect the fluence on longer time scales. This is demonstrated
in Fig. 2 by the two Monte-Carlo calculations, shown with the heavy
lines,  one of which includes a $1\mu{\rm G}$
magnetic field over the first $1{\rm kpc}$ propagation distance
to simulate the effect of a host galaxy magnetic field. 
It should be noted that the host galaxy
delay can be considerably shorter if the burst produces a significant 
amount of energy in protons with energies $\gtrsim 10^{21}{\rm eV}$, 
for which a host galaxy would introduce only a $\sim1{\rm h}$ delay.

Photons arriving with $\lesssim1{\rm day}$ delay are produced by cascades
initiated by pions created over the first 
$10{\rm Mpc}$ propagation distance of the protons. Most of these
pions are produced by protons with energy $\geq2\times10^{20}{\rm eV}$,
which lose $\sim1/2$ their energy over this distance.
The typical $>1{\rm TeV}$ fluence obtained on a day time scale, 
$\sim10^{-6}{\rm cm}^{-2}$ ($\sim10^{-5}{\rm erg\ cm}^{-2}$),
corresponds to $\sim2\%$ of the energy produced by the burst as
$\geq2\times10^{20}{\rm eV}$ protons.

\section{Conclusions}

Although a scenario in which CRs above $10^{20}{\rm eV}$
are produced by cosmological GRBs requires the arrival time of CRs to be 
delayed with respect to the $\gamma$-rays by more than $\sim50{\rm yr}$, 
we have shown that the delay of secondary, $0.01-100{\rm TeV}$ cascade
photons may be much smaller. A short delay is 
possible provided that a large fraction of the inter-galactic medium is 
occupied by large-scale magnetic ``voids'', regions of size
$\gtrsim5{\rm Mpc}$ and field weaker than $10^{-15}{\rm G}$.
For a field strength of $\sim4\times10^{-11}{\rm G}$ in the 
high-field inter-galactic regions, which would account for the observed 
galactic fields if it were frozen in the protogalactic plasma, the delay of 
CRs produced by a burst at a distance of $100{\rm Mpc}$ is $\sim100{\rm yr}$.
At the same time, the fluence of secondary $>1{\rm TeV}$ photons on a 
$1{\rm day}$ time scale would be $\sim10^{-6}{\rm cm}^{-2}$, provided
that $\sim80\%$ of the $20{\rm Mpc}$ region around the source is 
occupied by magnetic ``voids'' (The fluence is 
inversely proportional to the burst distance squared, since the photon time 
delay is independent of the burst distance).
The integral photon number
flux in the energy range $10{\rm GeV}<E<10{\rm TeV}$ is approximately 
$I(>E) \propto E^{-1.0}$. The flux at higher energies is very sensitive to
the infrared background intensity. If the intensity
is near the lower bound of current estimates, the flux extends
beyond $10{\rm TeV}$, approximately as $I(>E) \propto E^{-1.4}$.
If the intensity is close to its current upper bound, it would 
completely suppress the $>10{\rm TeV}$ flux from distances $\geq100{\rm Mpc}$.

A $3\sigma$ detection of $\geq1{\rm TeV}$ photons by current high-energy 
$\gamma$-ray experiments requires a fluence $\sim10^{-6} {\rm cm}^{-2}
\sqrt{t_{day}} E_{min,TeV}^{-1}$, where $t_{day}$ is the 
observation time measured in days and $E_{min,TeV}$ is the 
detector threshold energy in TeV (see \cite{CYGNUS} for CYGNUS, 
\cite{HEGRA} for HEGRA, \cite{CASA} for CASA-MIA, \cite{Whipple}
for Whipple). This fluence is close to that expected from a burst at
a distance of $\sim100{\rm Mpc}$. However, in a cosmological model,
the rate at which GRBs occur in a $100{\rm Mpc}$ sphere around us is low,
$\sim0.1{\rm yr}^{-1}$. A factor of $10$ increase in the sensitivity of TeV
detectors, as expected for example in the near future in the Whipple 
and HEGRA observatories,
would allow the detection of the delayed flux from
bursts occuring at distances up to $\sim300{\rm Mpc}$. The rate of such
bursts is $\sim2{\rm yr}^{-1}$. 

Lower threshold energy, space-based detectors such as EGRET
may also detect the delayed flux. At $10{\rm GeV}$, EGRET has an effective 
area of $\sim10^3\, {\rm cm}^2$. Thus, for the $>10{\rm GeV}$ fluence expected
from a burst at a $100{\rm Mpc}$ distance, $\sim10^{-4}{\rm cm}^{-2}$ in one
day, the probability that EGRET detects a $>10{\rm GeV}$ photon is
$\sim0.1$. This probability, although not negligible, is small.
Therefore, detection 
of the delayed flux would probably require next-generation GeV instruments, 
such as GLAST (\cite{GLAST}), that are expected to have order of magnitude 
better sensitivity
[It should be noted, however, that EGRET has detected
a $18{\rm GeV}$ photon from the direction of one of the strongest BATSE
bursts (second in fluence), with $\sim1.5{\rm hours}$ delay (\cite{GeV})].

\acknowledgements
We thanks B. Funk for valuable comments and information about HEGRA.
PC thanks the Institute for Advanced Study for its hospitality.
This research was partially supported by a W. M. Keck Foundation grant 
and NSF grants PHYPHY95-13835, PHY94-07194.

\newpage
\begin{figure}
\plotone{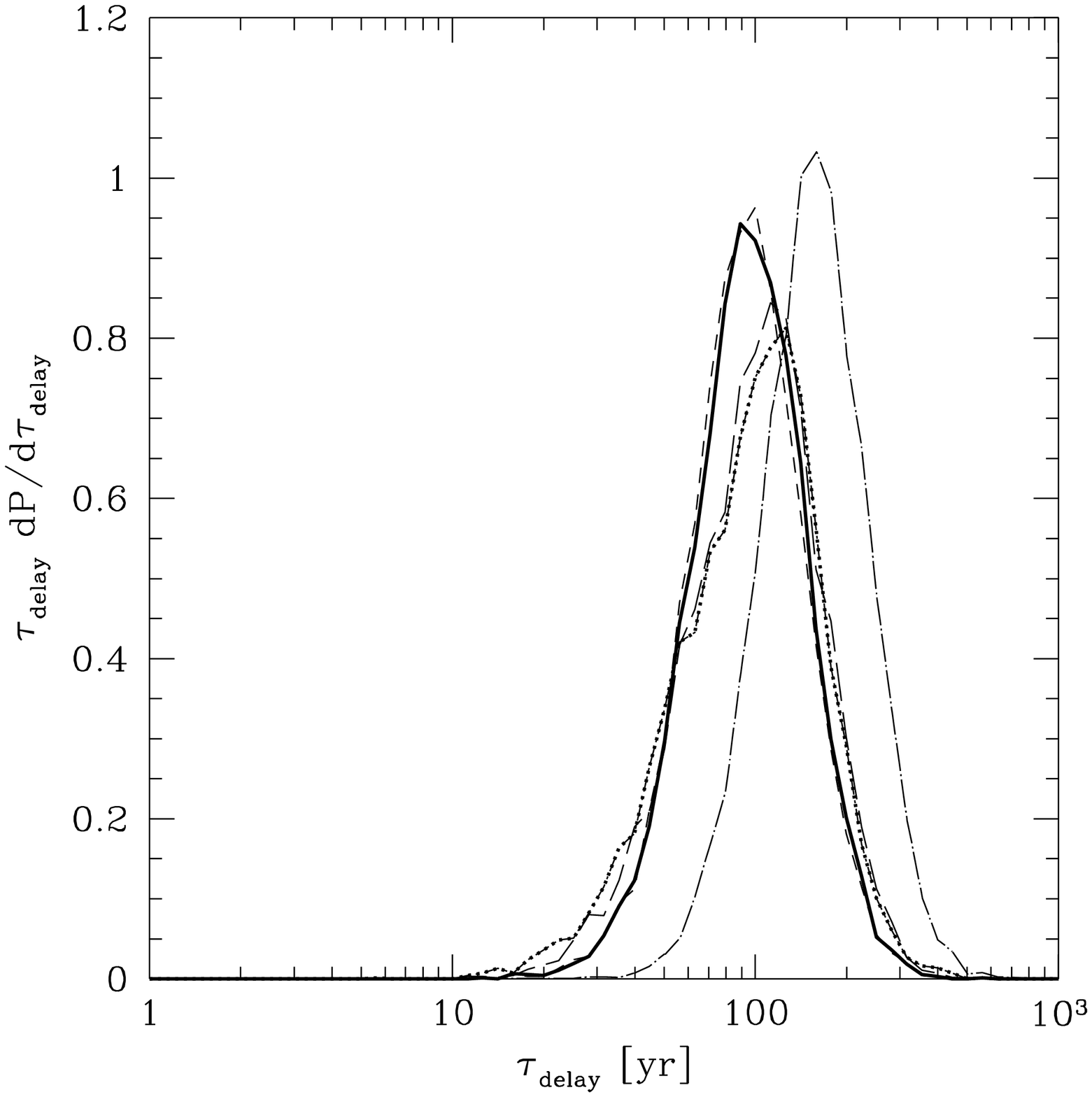}
\caption{
The distribution of arrival time delays of $>10^{20}{\rm eV}$ 
protons produced by
a burst at a distance of $100{\rm Mpc}$, for various IGMF structures
where a fraction $\sim0.2$ of the inter-galactic space is occupied by
$\sim1{\rm Mpc}$ regions of coherent $4\times10^{-11}{\rm G}$ field.
}
\label{fig1}
\end{figure}

\begin{figure}
\plotone{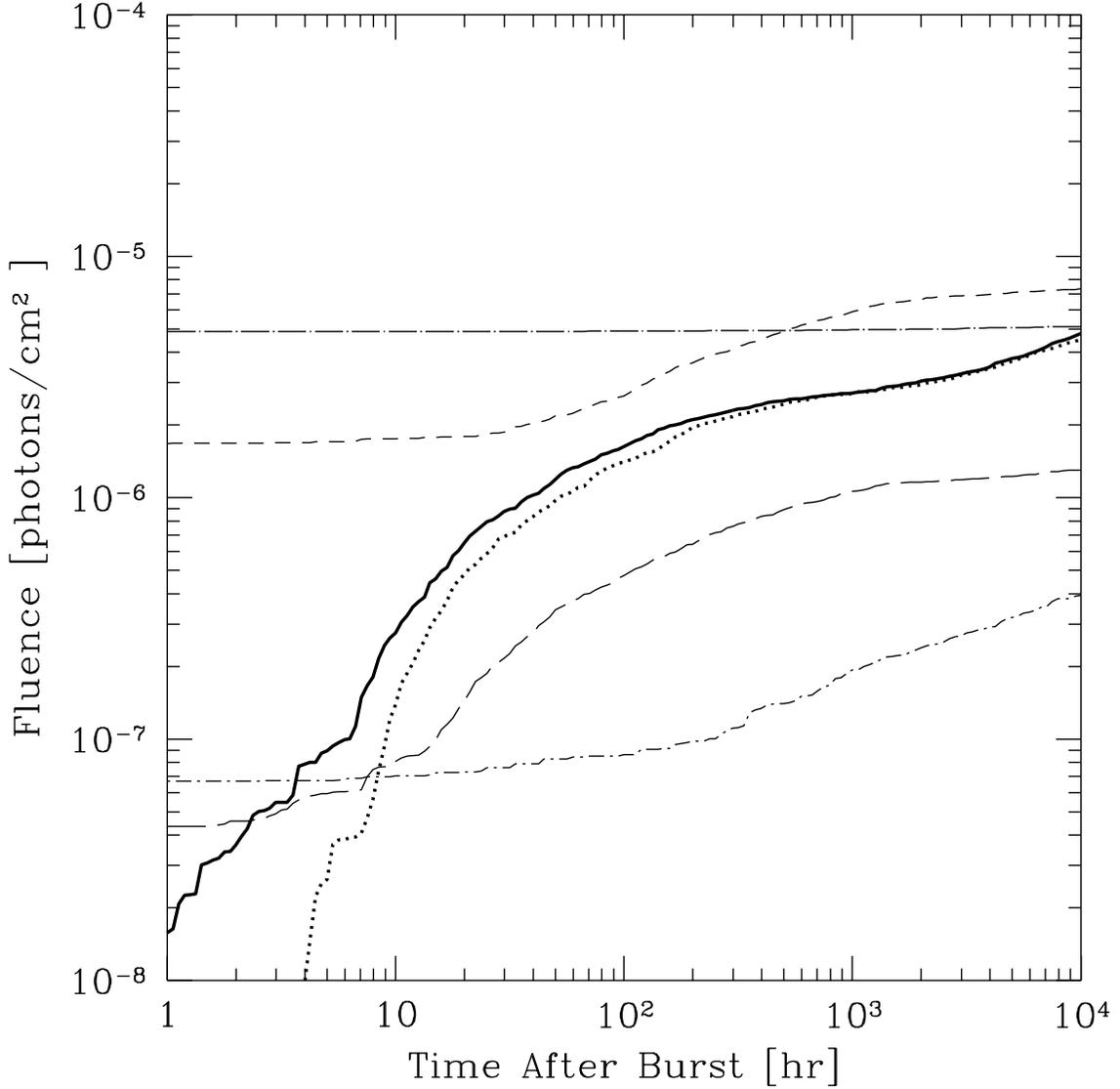}
\caption{
The fluence of $>1{\rm TeV}$ secondary photons, following a GRB producing
$10^{51}{\rm erg}$ as $>10^{20}{\rm eV}$ protons at a distance of
$100{\rm Mpc}$. The IGMF structures are the same as those used in Fig. 1,
and similar line-styles correspond to the same field structure in both
figures. The $heavy$ $dotted$ line gives the fluence obtained when a 
$1 \mu{\rm G}$ host-galaxy magnetic field is added to the 
field structure corresponding 
to the $heavy$ $solid$  line.
}
\label{fig2}
\end{figure}


\begin{thebibliography}{}

 \bibitem[Aharonian {\it et al.} 1994]{cas3}
Aharonian, F. A., Coppi, P. S., \& V\"olk, H. J. 1994, \apjl, 423, L5
 \bibitem[Alexandreas {\it et al.} 1991]{CYGNUS}
Alexandreas, D. E., {\it et al.} 1991, \apjl, 383, L53
 \bibitem[Bird {\it et al.} 1994]{Fly1} 
Bird, D. J., {\it et al.} 1994, \apj, 424, 491
 \bibitem[Bloom 1995]{GLAST}
Bloom, E. D. 1995, in {\it Proc. Workshop on TeV 
Gamma-Ray Astrophysics, Heidelberg, October 1994},
eds. Aharonian, F. A. and V\"olk, H. J.
 \bibitem[Cawley \& Weeks 1995]{Whipple}
Cawley, M. F., \& Weeks, T. C. 1995, to appear in {\it Experimental Astronomy}
 \bibitem[Cohen \& Piran 1994]{rate2} 
Cohen, E. \& Piran, T. 1995, \apjl, 444, L25 
 \bibitem[Cronin 1992]{huge} 
Cronin, J. W. 1992, Nucl. Phys. B (Proc. Suppl.) {28B}, 213
 \bibitem[Fishman \& Meegan 1995]{cos}
Fishman, G. J., \& Meegan, C. A. 1995, \araa, 33, 415
 \bibitem[Funk {\it et al.} 1995]{HEGRA}
Funk, B. {\it et al.} 1995, to appear in The proceedings for the Third 
Huntsville Symposium on Gamma-Ray Bursts, eds. C. Kouveliotou {\it et al.}
(AIP)
 \bibitem[Hurley {\it et al.} 1994]{GeV}
Hurley, K. {\it et al.} 1994, \nat, 372, 652
 \bibitem[Kosowsky \& Loeb 1996]{Avi}
Kosowsky, A., and Loeb A. 1996, submitted to the Ap. J., astro-ph/9601055
 \bibitem[Kronberg 1994]{Kron}
Kronberg, P. P. 1994, Rep. Prog. Phys. 57, 325
 \bibitem[Kulsrud {\it et al.} 1995]{Kulsrud}
Kulsrud, R., Cowley, S., Gruzinov, A., and Sudan, R., to appear in Phys. Rep.
 \bibitem[Kulsrud \& Anderson 1992]{Anderson}
Kulsrud, R. M., and Anderson, S. W. 1992, \apj, 396, 606
 \bibitem[Lee {\it et al.} 1995]{Olinto}
Lee, S., Olinto, A. V., \& Sigl, G. 1995, \apjl, 455, L21
 \bibitem[McKay {\it et al.} 1993]{CASA}
McKay, T. A., {\it et al.} 1993, \apj, 417, 742
 \bibitem[M\'esz\'aros 1995]{scen2} 
M\'esz\'aros, P. 1995, to appear in {\it Proc. 
17th Texas Conf. Relativistic Astrophysics} (NY Acad. Sci.)
 \bibitem[Milgrom \& Usov (1995)]{MU} 
Milgrom, M. \& Usov, V. 1995, \apjl, 449, L37
 \bibitem[Piran 1994]{scen1} 
Piran, T. 1994, in {\it Gamma-ray Bursts}, eds. G. Fishman
{\it et al.} (AIP 307, NY 1994)
 \bibitem[Plaga 1995]{Plaga}
Plaga, R. 1995, \nat, 374, 430
 \bibitem[Protheroe \& Johnson 1996]{cas1}
Protheroe, R. J., \& Johnson P. A. 1996, Astropar. Phys. in press (astro-ph/
9506119)
 \bibitem[Sironi {\it et al.} 1990]{Sironi}
Sironi, G., {\it et al} 1990, \apj, 357, 301
 \bibitem[Stecker {\it et al.} 1992]{Stecker}
Stecker, F.W., DeJager, O.C., \& Salamon, M.H. 1992, \apjl, 390, 49
 \bibitem[Vietri 1995]{Vietri} 
Vietri, M. 1995, \apj, 453, 883
 \bibitem[Waxman 1995a]{Wa} 
Waxman, E. 1995a, Phys. Rev. Lett. 75, 386
 \bibitem[Waxman 1995b]{Wb} 
Waxman, E. 1995b, \apjl, 452, L1
 \bibitem[Yoshida {\it et al.} 1995]{AGASA2}
Yoshida, S., {\it et al.} 1995, Astropar. Phys. {3}, 151

\end{thebibliography}
\end{document}